Highlights

**Global Climate Model Occultation Lightcurves Tested by August 2018 Ground-Based Stellar Occultation**

Sihe Chen,Eliot F. Young,Leslie A. Young,Tanguy Bertrand,François Forget,Yuk L. Yung

- We generated model lightcurves based on the state of Pluto's atmosphere simulated with a Global Climate Model (GCM).

- We used the Fourier optics method of Young (2012) to generate occultation lightcurves.

- Model occultation lightcurves are compared with ground-based observations of the Pluto stellar occultation event on 15 August 2018.

- The amplitude of the observed central flash is consistent with GCM simulations of Pluto in which a southern hemisphere $N_2$ ice band is present.

# Global Climate Model Occultation Lightcurves Tested by August 2018 Ground-Based Stellar Occultation

Sihe Chen[a,e,1]*, Eliot F. Young[b], Leslie A. Young[b], Tanguy Bertrand[d], François Forget[c] and Yuk L. Yung[a,f]

[a]*Division of Geological and Planetary Sciences, California Institute of Technology, Pasadena, CA 91125, USA*

[b]*Southwest Research Institute, Boulder, CO, USA 80302*

[c]*Laboratoire de Météorologie Dynamique, CNRS/UPMC (France)*

[d]*NASA/Ames Research Center, Moffett Field, California, USA*

[e]*School of Computing, National University of Singapore, Singapore 119077*

[f]*Jet Propulsion Laboratory, Pasadena, CA 91109*

---

**ARTICLE INFO**

**ABSTRACT**

Pluto's atmospheric profiles (temperature and pressure) have been studied for decades from stellar occultation lightcurves. In this paper, we look at recent Pluto Global Climate Model (GCM) results (3D temperature, pressure, and density fields) from Bertrand et al. (2020) and use the results to generate model observer's plane intensity fields (OPIF) and lightcurves by using a Fourier optics scheme to model light passing through Pluto's atmosphere (Young, 2012). This approach can accommodate arbitrary atmospheric structures and 3D distributions of haze. We compared the GCM model lightcurves with the lightcurves observed during the 15-AUG-2018 Pluto stellar occultation. We find that the climate scenario which best reproduces the observed data includes a $N_2$ ice mid latitude band in the southern hemisphere. We have also studied different haze and $P/T$ ratio profiles: the haze effectively reduces the central flash strength, and a lower $P/T$ ratio both reduces the central flash strength and incurs anomalies in the shoulders of the central flash.

---

## 1. Introduction

Pluto is one of four solar system objects on which the atmospheric column abundance of its primary gas components is known to be controlled by the vapor pressure of surface ices, the other three being Triton, Io, and Mars. Eris and other large trans-Neptunian objects are also likely

---

*Corresponding author

ORCID(s):0000-0002-0901-3428 (S. Chen); 0000-0001-8242-1076 (E.F. Young); 0000-0002-7547-3967

(L.A. Young); 0000-0002-2302-9776 (T. Bertrand); 0000-0002-3262-4366 (F. Forget); 0000-0002-4263-2562 (Y.L. Yung)

---

members of this group at some point in their orbits, although the presence of an atmosphere remains to be confirmed (Young et al., 2020).

Bertrand et al. (2020) have studied the surface-atmosphere interactions at Pluto and their impact on the near surface winds and the atmospheric general circulation by using a 3D GCM adapted to Pluto. Their model successfully reproduces one of the striking results of the New Horizons' volatile mapping campaign, with $CH_4$ ice predominantly located at high elevation while $N_2$ ice occupies low elevated terrains (Bertrand and Forget, 2016; Bertrand et al., 2019). Their model also points out the importance of Tombaugh Regio as the main $N_2$ reservoir that controls Pluto's column abundance.

On 15 August 2018, Pluto occulted the star UCAC4-341-187633. The event was observed by multiple sites located in Mexico, Texas, Wyoming, and along the U.S. east coast. More than three dozen lightcurves were obtained, including more than 10 lightcurves with the central flash features (Goldberg and Young, 2019). The primary purpose of this paper is to provide a technique that derives lightcurves from the GCM-simulated model atmospheres (Bertrand et al., 2020), and compare these lightcurves with observed ones.

The remaining parts of this paper describe the modeling methods (§2), details of the observation (§3), the comparison between model and observation results (§4), and conclusions (§5).

## 2. Methods

### 2.1. Simulation settings for the Pluto GCM

In order to predict the state of the atmosphere for the time of occultation, we employed the Laboratoire de Météorologie Dynamique (LMD) three-dimensional Pluto GCM. This model is designed to simulate the large-scale atmospheric dynamics on the whole planetary sphere and includes turbulence, radiative transfer, molecular conduction, convection, and phases changes for $N_2$, $CH_4$ and CO (in the atmosphere and at the surface). The detailed characteristics of the model are described in Forget et al. (2017) and Bertrand et al. (2020). We use the latest, most realistic, version of the model featuring:

- A digital elevation model (DEM) of the encounter hemisphere derived from New Horizons stereo imaging (Schenk et al., 2018). We use flat topography for most of the non-observed hemisphere as well as for the southern non-illuminated polar region.

- The presence of high-altitude perennial $CH_4$-rich deposits in the equatorial regions (known as the Bladed Terrain). These terrains are visible in the Tartarus Dorsa region (east of Sputnik Planitia) but their distinctive $CH_4$ absorption is seen in low resolution coverage of Pluto obtained during the New Horizons' approach phase, suggesting that they occur in patches further east along the equator (Olkin et al., 2017; Moore et al., 2018; Stern et al., 2020). In the GCM, we place a $CH_4$ ice reservoir at the locations of these terrains with a topography similar to that of the resolved Bladed Terrains in Tartarus Dorsa.

- The haze parameterization described in Bertrand and Forget (2017), which reproduces to first order the photolysis of $CH_4$ molecules in the upper atmosphere by Lyman-$\alpha$ UV radiation, the production of gaseous haze precursors, and their conversion into solid particles (using a simple formation scheme with a characteristic time for aerosol growth set to $10^7$ s). In the GCM, the haze particles are passive tracers with fixed uniform radii of 10 nm that only affects their sedimentation velocity. Although this parameterization is relatively simplified and not well validated, it remains reasonable to use it in this paper to estimate the vertical profile of haze density.

The simulations are performed using 32 longitude grid points and 24 latitude grid points (11.25x7.5, 150 km at the equator). In the vertical, 27 levels are typically used with most of the levels located in the first 15 km (the altitude of the first mid-layers are 5 m, 12 m, 25 m, 40 m) to ensure a suitable resolution in the lower troposphere and in the boundary layer. Above 10 km, the vertical resolution is about one scale height and the altitude of the top level is about 250 km, which corresponds to about 6 scale heights. As detailed in Forget et al. (2017) and Bertrand et al. (2020), the initial state of our GCM simulations (surface reservoirs of $N_2$, $CH_4$, and CO ices and the surface and subsurface temperatures) is the outcome of 30 million years of volatile ice evolution, with $N_2$ ice filling and flowing inside Sputnik Planitia, performed with the Pluto 2D volatile transport model, a reduced version of the GCM in which the 3D atmospheric transport and dynamics are replaced by a simple global mixing function of the volatile gases (Bertrand and Forget, 2016; Bertrand et al., 2018, 2019). The initial state used for the 3D GCM was selected among many possible results on the basis of their agreement with the observed pressure evolution and the distribution of surface ices in 2015 (Meza et al., 2019; Schmitt et al., 2017). We start the GCM simulation in 1984, in order to reach established methane and CO atmospheric cycles in equilibrium with the surface reservoir (Forget et al., 2017).

The simulations have been performed using an $N_2$ ice emissivity of 0.8 and an albedo of 0.7. The surface $N_2$ pressure simulated in the model is constrained by these values and reaches 1-1.2 Pa in 2015 as observed by New Horizons (Stern et al., 2015). The albedo and emissivity of the bare ground (volatile-free surface) are set to 0.1 and 1 respectively, which corresponds to a terrain covered by dark red materials such as the informally named Cthulhu Macula. $CH_4$ ice emissivity is fixed at 0.8. We use a $CH_4$ ice albedo of 0.5 for the equatorial deposits and an albedo of 0.75 for the polar $CH_4$ deposits based on albedo maps of Pluto (Buratti et al., 2017) and in order to give an atmospheric mixing ratio of $CH_4$ of about 0.3-0.6%.

The rest of the settings are similar to those in Bertrand and Forget (2017) and Bertrand et al. (2020).

## 2.2. Method for generating lightcurves

Young (2012) shows that the Fourier optics method works well for stellar occultation simulation for objects with thin atmospheres, in particular, Pluto. Moreover, the problem of Pluto's occultation falls well within the Fresnel regime. Trester (1999) introduces a modified aperture function $M(x_0, y_0)$ to calculate the observed electric field for Fresnel diffraction problems in the form of a Fourier transform (Equations 1-2).

$$M(x_0, y_0) = E_{inc}(x_0, y_0) Ap(x_0, y_0) \exp\left[\left(\frac{i\pi}{z\lambda}\right)(x_0^2 + y_0^2)\right] \qquad (1)$$

$$E_{obs}(x_1, y_1) = \mathcal{F}\{M(x_0, y_0)\} \qquad (2)$$

Where $x_0, y_0$ are the coordinates in the aperture plane, $x_1, y_1$ are the coordinates in the observer's plane, $E_{inc}, E_{obs}$ are the incident electric field onto the aperture and observer's plane respectively, $Ap$ is the aperture function, $\lambda$ is the wavelength, $z$ is the distance from aperture to observer's plane, and $\mathcal{F}$ is the Fourier transform operator. In this case, $z$ is the distance between Pluto and the Earth, which is equal to $4.97 \times 10^9$ kilometers at the time the occultation was recorded.

is restricted toHowever, a key restriction in using this form of Fourier transform is that spacing between points is restricted to $\sqrt{\lambda z / N}$, where $N$ is the number of grid points in each direction. Young (2012) indicates that this imposes some restrictions on generating Pluto occultation lightcurves: the value of $N$ required by Pluto occultation problems is too large for the Fourier transform to be practically calculated. To resolve this problem, one solution presented by Young (2012) is to use wavelengths longer than the visible wavelengths used in the actual observations.

In addition to the restriction on the grid spacing imposed, Fourier optics introduce diffraction patterns. However, since larger wavelengths are used in this paper, the patterns are bound to be nonphysical. To mitigate the diffraction patterns, a resampling process is introduced for the range of wavelengths, and the final result is the average value, normalized by the continuum. By the resampling process, the results of one wavelength are interpolated to the grid of another wavelength, since their corresponding grid spacings are different. In this paper, wavelengths of 5000 to 10000 $\mu m$ are used in steps of 1000 $\mu m$, and the results are interpolated to the smallest grid, corresponding to the wavelength of 5000 $\mu m$. This produces a Pluto-size field of view with a value of $N$ lower than 1024 and minimal diffraction patterns. Current Pluto occultation observations are typically sampled at 1 - 10 Hz, corresponding to sampling spacings of 1 - 10 km, too large by an order of magnitude to resolve visible wavelength diffraction features. We can therefore use Fourier techniques at longer wavelengths to solely reproduce the refractive effects of Pluto's atmosphere.

## 3. Observations

Pluto occulted a star (UCAC4-341-187633, $V = 13.0$) on 15-AUG-2018. This event was observed from sites in Mexico, Texas and along the U.S. east coast. Over three dozen useful lightcurves were obtained, including at least ten that show central flash features. In this paper we analyze two of these lightcurves with significant central flash features to constrain Pluto's surface pressure and haze opacity (Figure 1). The detailed characteristics of this stellar occultation are shown in Table 1.

| Date(UT) | RA_star (J2000) | DEC_star (J2000) | Vel. (km/s) | D (AU) | V_star |
|---|---|---|---|---|---|
| 2018-08-15 05:33:09.7 | 19h 22m 10.4687s | -21° 58' 49.020" | 19.33 | 32.7670 | 13.0 |

Table 1
Characteristics of the 15-AUG-2018 Stellar Occultation by Pluto.

### 3 .1. Occultation circumstances

The 15-AUG-2018 occultation was noteworthy for three reasons. First, the occulted star was bright, about four times brighter than Pluto in a V filter. Second, the prediction benefitted from the Gaia catalog and an accurate ephemeris for Pluto derived from the New Horizons flyby. Third, the shadow path crossed Mexico and the continental U.S., allowing the deployment of a picket fence of observers to sample the regions of the shadow in which a central flash was expected. Our closest portable telescope provided a central flash that was higher than the out-of-event stellar baseline, the first time that had ever been observed during a Pluto stellar occultation.

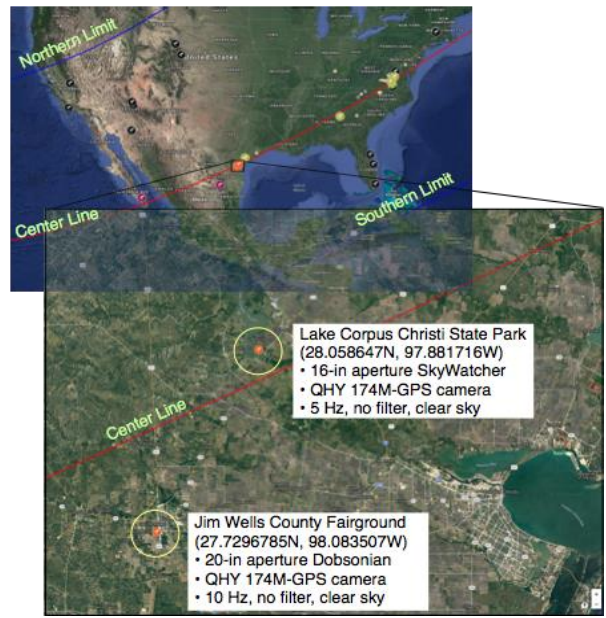

Figure 1: The upper map shows the predicted shadow path across Mexico and the U.S., with the centerline (red) and the region near Alice, TX where several telescopes were deployed (orange rectangle). The lower panel is a close-up of the orange rectangle, showing the locations of portable telescopes at Lake Corpus Christi State Park (LCC) and Jim Wells County Fairground (JWF).

### 3.2. Extracting the lightcurves

We analyzed time-tagged images from LCC (5 Hz) and JWF (10 Hz) for tens of minutes before and after the occultation. We identified stars in each frame with astrometry.net software (Lang et al., 2010), then used a form of aperture photometry to estimate the brightness of Pluto and the occulted star relative to other field stars. We extracted the lightcurve using a PSF-fitting algorithm as well, but the signal-to-noise ratio (SNR) was better with the aperture method than the PSF scheme (16.1 vs. 12.3 per point, and 81.9 vs. 44.2 per 50 km in comparison to the LCC lightcurve). The lightcurves are normalized so that the out-of-event stellar baseline has amplitude of 1.

## 4. Results

### 4.1. $N_2$ surface ice distribution

Bertrand et al. (2020) presented different possible scenarios for $N_2$ surface ice distribution. As observed by New Horizons, there are $N_2$ ice deposits in the low-elevation terrains of the northern mid-latitudes (Schmitt et al., 2017; Protopapa et al., 2017; Howard et al., 2017). Since the southern hemisphere was not observed, it is possible that there is an extra mid-latitudinal band of $N_2$ ice deposits as well. This would remain consistent with the increase of atmospheric pressure observed during the 1990-2015 period. The associated $N_2$ north-south sublimation-condensation flow can significantly influence the haze distribution. Effectively, if the band exists, the $N_2$ flow pushes the haze particles towards the south pole, as shown on the right panel of Figure 2. An example of the haze distribution is shown in Figure 2. From the figure, we observe that the haze distributions at low altitudes are apparently more concentrated towards the south pole in case with a southern hemisphere $N_2$ ice band. In this paper, we investigated the two different scenarios with GCM model lightcurves, with or without an extra mid-latitudinal band of $N_2$ ice deposits.

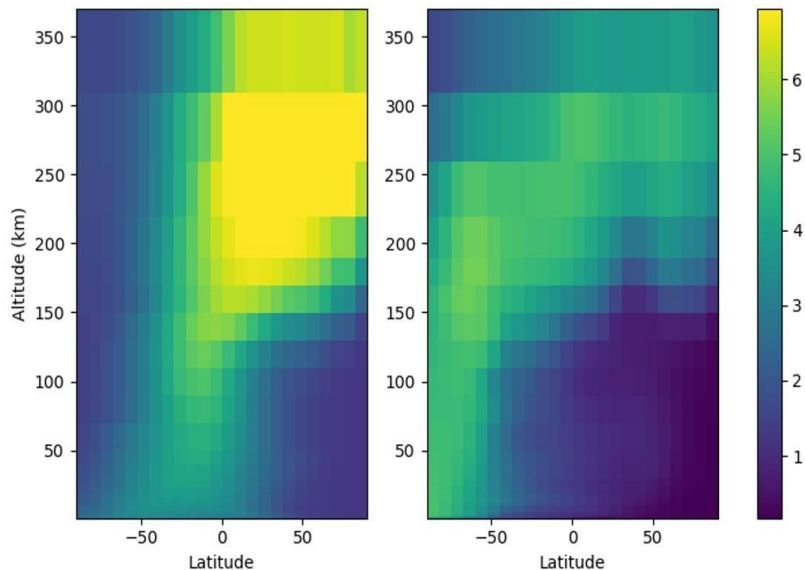

Figure 2: Haze distributions for cases without (left) or with (right) a southern hemisphere $N_2$ ice band. The color stands for mass fraction of haze particles, in unit of $10^{-6}$ kg haze/kg air.

**4.2. The GCM model lightcurves**

For the Fourier optics approach, an aperture screen needs to be constructed from the GCM data first. It is calculated by integrating along chords corresponding to the discrete points in the aperture screen, with the values on the chords interpolated from the GCM grid. The refractivity used for this integration process is determined with a two-term Sellmeier formula (Griesmann and Burnett, 1999). The temperature and pressure profiles are used to calculate the refractivity field; moreover, the wavelengths used for refractivity calculation are visible wavelengths, rather than the wavelength used in the Fourier optics method.

Moreover, the haze opacity is calculated from the mass fractions of the haze particles provided in the GCM data. The haze particles are assumed to have a spherical shape. The optical depth is calculated with Equation 3.

$$\tau = \frac{3Q_{ext}M_a}{4\rho_a r_a} \tag{3}$$

where $M_a$ is the column mass of haze (g/cm$^2$), $\rho_a$ is the density of haze particles (1g/cm$^3$), $r_a$ is the radius of haze particles (10 nm), and extinction coefficient, $Q_{ext} = 0.007$ (Bertrand and Forget, 2017). The global average normal optical depth $\tau_N$ is consistent with Gladstone et al. (2016) and is 0.0022 and 0.0028 with or without the N$_2$ ice band at the time of the occultation, respectively.

The longitude and latitude of the center of the Pluto as seen from the earth at the time of the occultation event are 70.90° W and 54.28° N. The GCM grid is accordingly rotated before the calculation.

The model lightcurves and observer's plane intensity field (OPIF) are generated from the aperture screen with the Fourier optics method introduced in §2 (Figure 3). Its intensity is normalized so that the out-of-event intensity is equal to 1. Differences are observed for the magnitudes of the central peak and the shapes of the ingress and egress shoulders. For example, in the egress shoulder of the no-band case, there is a position where the third derivative turns to zero (at 1000 km). One more noteworthy pattern is the asymmetry in OPIF: for both cases, the asymmetrical patterns near ($x$=1000 km, $y$=1000 km) are apparent. Asymmetrical structures are essentially enhancements in certain locations and diminutions in other locations, and these structures can only be observed when the model OPIF is derived from GCM data. When crossed by the track, the asymmetry could be observed in the lightcurves. For example, the ingress and egress shoulders are observably different in lightcurves for the no-band model (Figure 3).

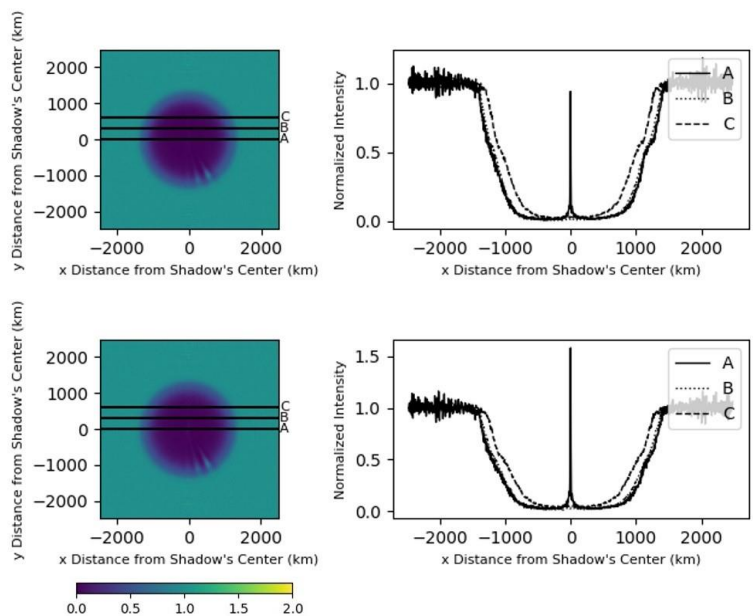

Figure 3: Model OPIF and lightcurves generated for the GCM data, without (upper figures) or with (lower figures) the southern hemisphere $N_2$ ice band. The center of the OPIF corresponds to 70.90° W and 54.28° N on Pluto, and projections of both north and south poles lie on the y axis. The corresponding tracks of the lightcurves are marked in the OPIF, located at the central track (A), 300 km away from the central track (B), and 600 km away from the central track (C), respectively.

    To illustrate the ability of this technique to capture the 3D features, results for different viewing angles are obtained as well. In Figure 4, it is shown that the asymmetrical patterns are seen in different positions, have different strengths, or can be missing. The results shown in Figure 4 suggest that subtle asymmetrical structures in the lightcurves can provide extra information. Although the lightcurves recorded in the August 2018 event did not cross the asymmetrical structures predicted by GCM, we expect that some future events could provide observations on the asymmetrical patterns.

GCM Occultation Tested by Observation

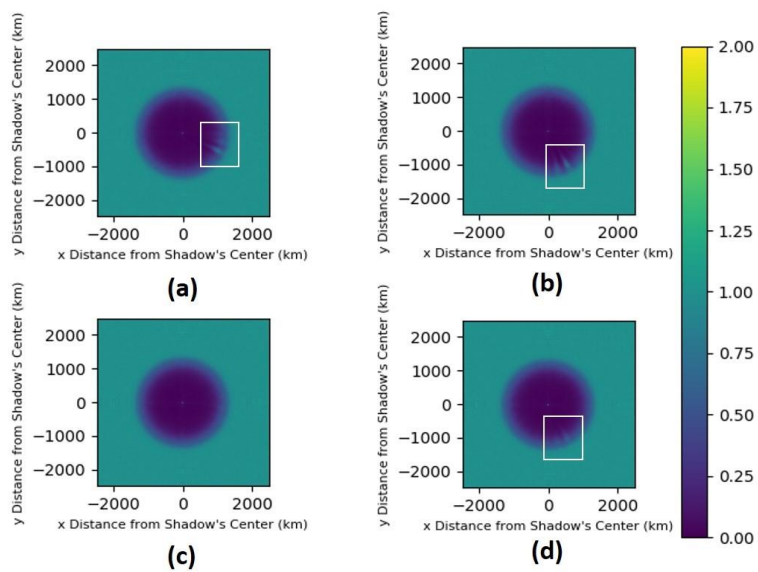

Figure 4: Model OPIFs generated for different sub-earth longitudes and latitudes: (a) 160.90° W and 54.28° N, (b) 109.10° E and 54.28° N, (c) 0° E and 0° N, (d) 35.72° W and 54.28° N. (a), (b), and (d) are 90° separated from the original viewing point along meridians or parallels. The white frames mark the asymmetrical patterns. The model with the southern hemisphere $N_2$ ice band is used here, and projections of both north and south poles lie on the y axis.

## 4.3. Comparisons between observed and model lightcurves

The comparison between the observation and model is of great interest since it reveals the validity and sensitivity of the model. When generating the model lightcurves, by taking into consideration the orbital geometries and axis tilts of the two bodies at the time of occultation, tracks making -35.1° with respect to the Pluto's equator would be appropriate.

Model lightcurves are extracted from the OPIF and compared with the observed lightcurves (Figure 5), assuming that the track crosses the center of Pluto's shadow.

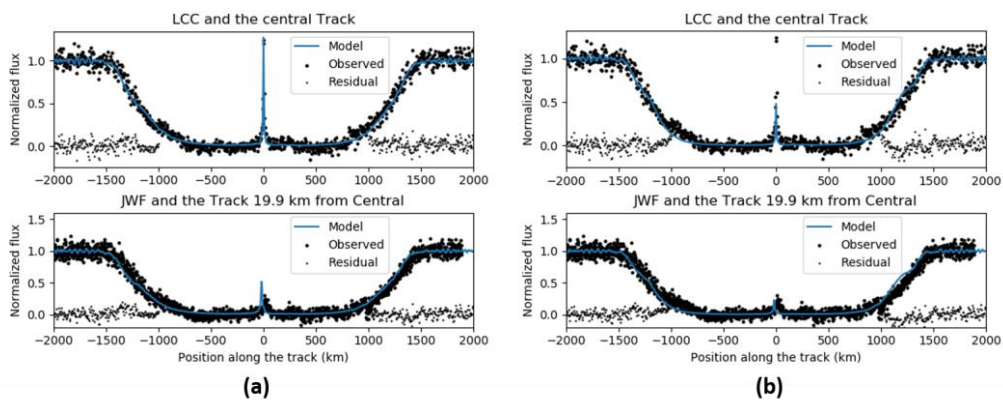

Figure 5: Comparisons between observed and GCM model lightcurves, for cases (a) with or (b) without the southern hemisphere $N_2$ ice band. "Position along the track" stands for the position on the observer's plane along the track that the observer travels along (aka "downtrack" position).

A closer look at the ingress and egress shoulders are shown in Figure 6. It can be concluded from the figures that the lightcurves predicted by the both cases agree well with the observed data at the shoulder region. The sum of squared residuals in the shoulder regions is 0.68 for the case without band, and slightly lower, 0.60 for the case with band. However, the difference is too small to exclude the possibility of the no-band case. The difference between normalized fluxes of the two cases in the shoulder region is no larger than 0.04, and the average is 0.01. For different tracks, the differences are of the same order. Hence, the anomaly could not be identified when we have an SNR of 16.

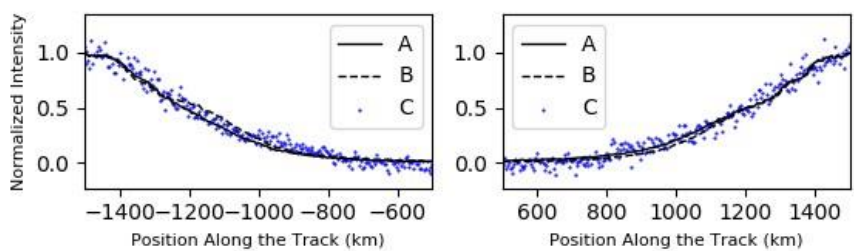

Figure 6: Comparisons between observed and GCM model lightcurves, with the GCM model lightcurves (A) with $N_2$ ice deposit band in southern hemisphere, (B) without $N_2$ ice deposit band in southern hemisphere, and (C) the data points. "Position along the track" stands for the position on the observer's plane along the track that the observer travels along (aka "downtrack" position).

Another important investigation is made on the central flash strength. Figure 5b shows that for the no-band case, the normalized central flash strength is 0.48, while Figure 5a shows that with band the central flash strength is 1.27. The lower central flash strength in the no-band case can be attributed to the light being refracted to the center passing through haze with higher optical depth. In comparison, the observed central flash strength is 1.24, and the case with band provides better match with the observation. Moreover, in Figure 5b we observe a rise in the residuals at about 1000 km from the center. The comparison results for the central flash strength indicate a higher probability that the band of N$_2$ ice deposit in the southern hemisphere exists. If we adopt the case with the band, across both LCC and JWF tracks, the residuals are kept at low level. The mean of squared residuals in the lightcurves is 0.003 for LCC, and 0.006 for JWF. In addition, Arimatsu et al. (2020) indicates a pressure drop on Pluto with an occultation event in 2019, agreeing with the prediction of the case with band.

### 4.4. Sensitivities to haze optical depth and $P/T$ ratio profiles

As shown and discussed in Figure 3 and §4.2, the no-band case shows some asymmetrical patterns in the egress shoulder. By manipulating the temperature, pressure, and haze profiles, we can find the effect of each quantity on the generated lightcurve and source of the asymmetries. Two adjusments are done on the GCM: based on the no-band GCM data, we either eliminate the haze globally, or scale the $P/T$ ratio and reduce it by 25% globally.

The comparison of the lightcurves generated by the original no-band GCM and the two variations are shown in Figure 7. The comparison in the central flash region is straightforward: without haze, the central flash strength is increased, and reducing $P/T$ ratio globally decreases the central flash strength. The effect of haze can be easily explained, since haze absorbs light and brings down the intensity. The effect of scaling $P/T$ ratio, however, is more complicated. Essentially, the central flash is generated by the light refracted to reach the center, and the atmosphere closer to the surface of Pluto generally refracts the light more strongly. If $P/T$ ratio is reduced, the refractivity is proportionally reduced, and the central flash has to come from lower altitudes, where the haze opacity is larger. Therefore, the central flash strength is reduced.

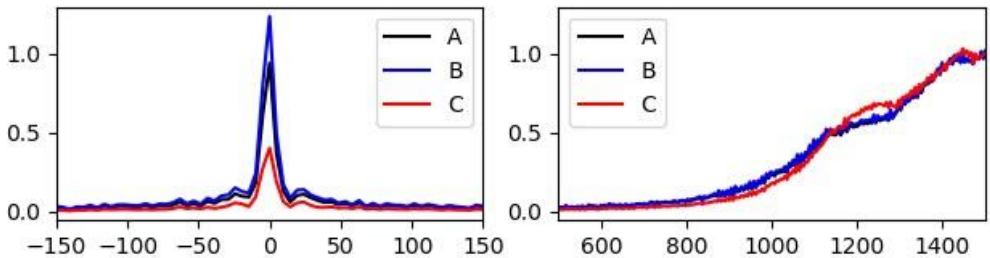

Figure 7: Model lightcurves of the central track when we use: (A) the complete model, (B) the model without haze, (C) the model with $P/T$ ratio reduced by 25% globally. The comparisons for the central flash region (left) and the egress shoulder (right) are shown. The model lightcurves are generated by the case with no southern-hemisphere N$_2$ band present.

Figure 7 also shows the comparisons in the shoulder. Whether haze is considered makes minimal difference to the egress shoulder. This explains the reason why the different haze distributions of the cases with or without band generate no distinguishable differences in the shoulder region, as shown in Figure 6. On the other hand, the reduced $P/T$ ratio enhances the asymmetry, and even a turning point appears at 1200 km from the center. This suggests that this asymmetry is induced temperature enhancement.when the $P/T$ ratio near the egress shoulder is lower than the ingress shoulder, such as a local temperature enhancement.

## 5. Conclusions

Model OPIF and lightcurves are generated with GCM (Bertrand et al., 2020) and Fourier optics method (Young, 2012). The results are compared with observed lightcurves in the 15-AUG-2018 Pluto stellar occultation event, and it is shown that good agreements with observed data are obtained when the GCM model has a $N_2$ ice band in the southern hemisphere. The effect of different haze and $P/T$ ratio profiles are also studied. While the haze reduces the central flash strength, a lower $P/T$ ratio both reduces the central flash strength and incurs anomalies in the shoulders.

**Acknowledgements:** This work is supported in part by NASA 000329-P2232440 to Caltech, and by NSF 1616115 and NASA SSO 80NSSC19K0824 to SwRI.